\documentclass[epsfig,12pt,onecolumn]{article}
\usepackage{amsfonts}
\usepackage{amssymb}
\usepackage{amsmath}
\usepackage{multicol}
\usepackage{graphicx}
\usepackage{float}
\usepackage{caption}

\input{tcilatex}
\makeatletter \@addtoreset{equation}{section}

\def \be{\begin{equation}}
\def \ee{\end{equation}}
\def \bea{\begin{eqnarray}}
\def \eea{\end{eqnarray}}

\newcommand{\nc}{\newcommand}
\nc{\al}{\alpha} \nc{\bib}{\bibitem} \nc{\la}{\lambda}
\nc{\C}{\mbox{\hspace{1.24mm}\rule{0.2mm}{2.5mm}\hspace{-2.7mm} C}}
\nc{\R}{\mbox{\hspace{.04mm}\rule{0.2mm}{2.8mm}\hspace{-1.5mm} R}}

\begin{document}

\title{\textbf{Holographic dark energy satisfying the energy conditions in
Lovelock gravity}}
\author{M. Bousder$^{1}$\thanks{%
mostafa.bousder@um5.ac.ma}, E. Salmani$^{2,3}$, A. El Fatimy$^{4,5}$ and H.
Ez-Zahraouy$^{2,3}$ \\
$^{1}${\small Faculty of Sciences, Mohammed V University in Rabat, Morocco}\\
$^{2}${\small Laboratory of Condensend Matter and Interdisplinary Sciences,
Department of physics,}\\
\ {\small Faculty of Sciences, Mohammed V University in Rabat, Morocco}\\
$^{3}${\small CNRST Labeled Research Unit (URL-CNRST), Morocco}\\
$^{4}${\small Central European Institute od Technology,}\\
\ {\small CEITEC BUT, Purky\v{n}ova 656/123, 61200 Brno, Czech Republic}\\
$^{5}${\small Departement of Physics, Universit\'{e} Mohammed VI
Polytechnique, Ben Guerir 43150, Morocco}}
\maketitle

\begin{abstract}
In this paper, we show that the holographic dark energy density hides in the
solutions of Lovelock gravity for black holes. Using the obtained mass and
temperature we find density equations. We propose a physical interpretation
of the rescaled Lovelock couplings as a topological mass that describes the
Lovelock branch. In addition to this, we present new solutions that satisfy
the energy conditions according to the Lovelock coupling and the horizon
curvatures. This work can be extended to the equation of the state $\omega
_{\Lambda }$ of dark energy in third-order Lovelock gravity. We show that
the value $"-1"$ represents a stable equilibrium of $\omega _{\Lambda }$.
\end{abstract}

\section{Introduction}

Recently, in the context of Lovelock gravity, holographic implications\ \cite%
{ZL} and boundary conditions \cite{ZL1} are obtained. Lovelock gravity
represents a large class of gravitational theories that have several
important properties. It is possible to show the Smarr formula and an
extended first law for the Lovelock black holes, which gives the mass in
terms of geometrical quantities \cite{INi}. It should be noted that the
Lovelock-Lanczos terms can provide a non-trivial contribution to the
Einstein field equations in four dimensions \cite{INI2}. The equations of
motion of Lovelock gravity are obtained by varying the action of the
Lovelock gravity \cite{ANN1,ANN2,ANN3}. It is worthwhile to mention that the
black hole solutions are also obtained in Lovelock gravity coupled to
Born-Infeld with non-constant curvatures \cite{IN3}. These solutions can be
considered within the presence of a perfect fluid for a linear equation of
state \cite{IN4}: $p=\omega \left( \rho -4B\right) $. The most extensively
researches are done on the second-order curvature corrections
(Einstein-Gauss-Bonnet gravity) \cite{GB00,GB0,GB1,GB2}. There are several
applications of third-order Lovelock gravity as a new class of black hole
solutions with supplementary conditions on their Weyl tensors \cite{IN2}.%
\textrm{\ }The application of Lovelock theory can be extended to the 4D
charged black holes with ModMax nonlinear electrodynamics \cite{INM} and
evolving wormhole configurations \cite{N3}. In \cite{Z1} they have treated
the cosmological constant as thermodynamic pressure and its conjugate as a
thermodynamic volume with different normalized sectional curvature.
Furthermore, the relationship between black holes and dark energy is studied
in a series of papers \cite{N1, N2,NX1}. In addition to this, the
energy-momentum tensor $T_{\mu \nu }$ for the matter field satisfies the
general case energy conditions \cite{EC}. By taking into account any null
vector $k^{\mu }$ and for any timelike vector $v^{\mu }$: (a) Null energy
condition (NEC) is $T_{\mu \nu }k^{\mu }k^{\nu }\geq 0$. (b) Weak energy
condition (WEC): $T_{\mu \nu }v^{\mu }v^{\nu }\geq 0$. (c) Strong energy
condition (SEC): $\left( T_{\mu \nu }-\frac{1}{D-2}Tg_{\mu \nu }\right)
v^{\mu }v^{\nu }\geq 0$.\ (d) Dominant energy condition (DEC): $T_{\mu \nu
}v^{\mu }v^{\nu }\geq 0$ and $J_{\mu }J^{\mu }\leq 0$ with $J_{\mu }=-T_{%
\text{ \ }\nu }^{\mu }v^{\nu }$. In this set-up, the equivalent expressions
of the energy conditions for perfect fluid for $T_{\mu \nu }=\left( \rho
+p\right) u_{\mu }u_{\nu }+pg_{\mu \nu }$ with $u_{\mu }$ is the fluid
four-velocity \cite{EC,ED} are given by:\newline
$\bullet $ NEC: $\rho +P\geq 0$.\newline
$\bullet $ WEC: $\rho +P\geq 0$ and $\rho \geq 0$.\newline
$\bullet $ SEC: $\rho +P\geq 0$ and $\left( D-3\right) \rho +\left(
D-1\right) P\geq 0$.\newline
$\bullet $ DEC: $-\rho \leq P\leq \rho $ and $\rho \geq 0$.\newline
The energy conditions have not yet been dealt with in the framework of
Lovelock black holes with dark energy (DE). Nevertheless, traversable
wormholes satisfying just the WEC in third-order Lovelock gravity \cite%
{T1,N4,N5,N6}.\newline
More recently there has been a lot of effort to construct more models of
holographic DE \cite{D9,D10,DP1}, which are characterized by an equation of
state and energy density with a radius of the cosmological horizon \cite{DP2}%
. While the Barrow entropy of black hole \cite{DD8} plays a very important
role in the description of holographic dark energy \cite{HDH,DD10}. This is
in good agreement with the holographic principle \cite{D11,D12}. The
holographic inflation \cite{HI} is one of the existing models for quantum
gravitational implications on the black hole horizons. Also, the brane
inflation has been explored in \cite{L7,L8}. Moreover, black holes can also
have formed in the early Universe \cite{L9,L10,L11}. Inspired by the recent
studies on the Lovelock gravity in the spherically symmetric formalism%
\textbf{, }we propose a new method to determine the energy conditions of
dark energy in the presence of black holes.\newline
The paper is planned as follows: In section 2, we study the density of the
holographic dark energy in D-dimensional Lovelock gravity. In Section 3, we
review the thermodynamic aspect and the spherically symmetric solutions in
third-order Lovelock gravity. In section 4, we study different densities
describing the black hole mass and energy conditions. Also, we analyze the
equation of state parameter of dark energy. Subsequently, in Section 5, we
discuss the solutions with the energy conditions using the Hawking
temperature. We discuss the relationship between the Hubble parameter and
the rescaled Lovelock coupling. We conclude our findings in Section 6.%
\textbf{\ }Throughout this article, we use units $G=c=\hbar =k_{B}=1$.

\section{Equation of state in Lovelock gravity}

The Lagrangian of a Lovelock gravity \cite{Z13} of extended Euler densities
in D spacetime dimensions is given by%
\begin{equation}
\mathcal{L}=\frac{1}{16\pi G_{N}}\sum_{k=0}^{K=\left[ \frac{D-1}{2}\right] }%
\frac{\alpha _{k}}{2^{k}}\delta _{c_{1}d_{1}\cdots
c_{k}d_{k}}^{a_{1}b_{1}\cdots a_{k}b_{k}}R_{a_{1}b_{1}}^{\text{ \ \ \ \ }%
c_{1}d_{1}}\ldots R_{a_{k}b_{k}}^{\text{ \ \ \ \ }c_{k}d_{k}},  \label{a1}
\end{equation}%
where $\alpha _{k}$ are the Lovelock coupling constants, $\delta
_{c_{1}d_{1}\cdots c_{k}d_{k}}^{a_{1}b_{1}\cdots a_{k}b_{k}}$\ is the
generalized totally antisymmetric Kronecker delta and $R_{a_{1}b_{1}}^{\text{
\ \ \ \ }c_{1}d_{1}}$ is the generalized Riemann tensor. The brackets $\left[
\cdots \right] $ denote taking the integer part, and $\alpha _{2}\mathcal{\ }
$and $\alpha _{3}$ are respectively the second (Gauss-Bonnet coupling) and
the third Lovelock coefficients. The zero-order Lovelock invariant is $%
\alpha _{0}L_{0}=-2\Lambda =\frac{\left( D-1\right) \left( D-2\right) }{L^{2}%
}$ with $\Lambda $ being the cosmological constant and $L$ is the AdS
radius. We introduce the effective thermodynamic pressure \cite{CQG2} as%
\begin{equation}
P=\frac{\alpha _{0}}{16\pi G_{N}}=-\frac{\Lambda }{8\pi G_{N}}.  \label{a2}
\end{equation}%
The first-order Lovelock invariant is $\alpha _{1}L_{1}=R$ is the
Einstein-Hilbert Lagrangian. The corresponding solution is given by the
metric function \cite{CQG}:
\begin{equation}
f(r)=\kappa +r^{2}\left( K\alpha \right) ^{-\frac{1}{K-1}}\left( 1-\left( 1+%
\frac{m\left( r\right) -\alpha _{0}}{\alpha \left( K\alpha \right) ^{-\frac{K%
}{K-1}}}\right) ^{1/K}\right) ,  \label{a7}
\end{equation}%
with $\kappa $ is the horizon curvatures with $\kappa =\left\{ 0,\pm
1\right\} $ for\ $\Lambda <0$ and $\kappa =+1$ for\ $\Lambda \geq 0$. Thus
we can uniformly rewrite the Lovelock equations as the Arnowitt-Deser-Misner
(ADM) mass $M$ of black hole \cite{NPB1}:%
\begin{equation}
m(r)=\frac{16\pi G_{N}M}{\left( D-2\right) \Sigma _{D-2}r^{D-1}}=\frac{%
\alpha _{0}}{\left( D-1\right) \left( D-2\right) }+\sum_{k=0}^{K}\alpha
_{k}\left( \frac{\kappa -f(r)}{r^{2}}\right)
^{k}\prod\limits_{l=3}^{2k}\left( D-l\right) ,  \label{a5}
\end{equation}%
where $\Sigma _{D-2}$ is the area of a radius $(D-2)$-dimensional unit
sphere: $\Sigma _{D-2}=\frac{2\pi ^{\frac{\left( D-1\right) }{2}}}{\Gamma
\left( \frac{D-1}{2}\right) }.$ The $\left( D-1\right) $-dimensional volume
of a Euclidean ball of radius $R_{H}$ is%
\begin{equation}
V_{D-1}=\frac{2\pi ^{\frac{\left( D-1\right) }{2}}}{\left( D-1\right) \Gamma
\left( \frac{D-1}{2}\right) }R_{H}^{D-1}=\frac{\Sigma _{D-2}R_{H}^{D-1}}{%
\left( D-1\right) },  \label{a6}
\end{equation}%
Here, $R_{H}$\ denotes the radius of the black hole is one of the roots of
the metric function Eq. (\ref{a7}). Using the D-dimensional static solutions
in the Lovelock gravity for $\alpha _{k}=\alpha \left( K\alpha \right) ^{-%
\frac{K-k}{K-1}}\left(
\begin{array}{c}
K \\
k%
\end{array}%
\right) $, $2\leq k\leq K$. The event horizon in spacetime can be located by
solving the metric equation: $f(r)=0$, which yields%
\begin{equation}
\frac{m\left( r\right) }{\alpha \left( K\alpha \right) ^{-\frac{K}{K-1}}}=-1+%
\frac{\alpha _{0}}{\alpha \left( K\alpha \right) ^{-\frac{K}{K-1}}}+\left[
1+\kappa \frac{\left( K\alpha \right) ^{\frac{1}{K-1}}}{r^{2}}\right] ^{K}.
\label{a8}
\end{equation}%
One should note that in the event horizon $R_{H}$ we have $m(R_{H})=\frac{%
16\pi G_{N}M}{\left( D-2\right) \Sigma _{D-2}R_{H}^{D-1}}$, so that one
obtains the following solution%
\begin{equation}
\frac{16\pi G_{N}\left( K\alpha \right) ^{\frac{K}{K-1}}\rho _{BH}}{\alpha
\left( D-2\right) \left( D-1\right) }=-1+\frac{\alpha _{0}}{\alpha }\left(
K\alpha \right) ^{\frac{K}{K-1}}+\left[ 1+\kappa \frac{\left( K\alpha
\right) ^{\frac{1}{K-1}}}{R_{H}^{2}}\right] ^{K}.  \label{a10}
\end{equation}%
Here $\rho _{BH}=M/V_{D-1}$ is the $\left( D-1\right) $-dimensional black
hole density. We would like to mention here that from \cite{L13}, there is a
connection between black holes and holographic dark energy. Also, the
holographic vacuum energy in a volume should not exceed the energy of a
black hole of the same size \cite{L14}. Using the binomial theorem
\begin{equation}
\left[ 1+\frac{\kappa \left( K\alpha \right) ^{\frac{1}{K-1}}}{r^{2}}\right]
^{K}=\sum_{k=0}^{K}\left(
\begin{array}{c}
K \\
k%
\end{array}%
\right) \left( \frac{\kappa \left( K\alpha \right) ^{\frac{1}{K-1}}}{r^{2}}%
\right) ^{k},  \label{bn}
\end{equation}%
and the density of the holographic dark energy \cite{D9,HDH,HDE2}: $\rho
_{\Lambda }\equiv \frac{3}{16\pi G_{N}R_{H}^{2}}.$ By considering a specific
pressure of solutions corresponding to the choice of the equation of state
(EoS) parameter of dark energy $\omega _{\Lambda }=-$ $P/\rho _{\Lambda }$.
From Eqs. (\ref{a10})-(\ref{bn}) we obtain%
\begin{equation}
\frac{\rho _{BH}}{\left( D-2\right) \left( D-1\right) }=P+\frac{\kappa }{3}%
\rho _{\Lambda }+\alpha \left( K\alpha \right) ^{-\frac{K}{K-1}%
}\sum_{k=2}^{K}\left(
\begin{array}{c}
K \\
k%
\end{array}%
\right) \frac{\left( \kappa \left( K\alpha \right) ^{\frac{1}{K-1}}\right)
^{k}}{16\pi G_{N}R_{H}^{2k}}.  \label{a11}
\end{equation}%
Taking $L=H^{-1}$ as the effective Hubble horizon, thus the Friedmann
equation $HR_{H}=1$ \cite{D9} and using Eq. (\ref{a12}) $\rho _{\Lambda }=%
\frac{3}{16\pi G_{N}R_{H}^{2}}.$ It is easy to show that%
\begin{equation}
\omega _{\Lambda }=\frac{\kappa }{3}-\frac{16\pi G_{N}\Gamma \left( \frac{D-1%
}{2}\right) }{3\left( D-2\right) 2\pi ^{\frac{\left( D-1\right) }{2}}}%
MH^{D-3}+\frac{\alpha }{3}\sum_{k=2}^{K}\left(
\begin{array}{c}
K \\
k%
\end{array}%
\right) \kappa ^{k}\left( K\alpha \right) ^{\frac{k-K}{K-1}}H^{2k-2}
\label{a12}
\end{equation}%
For a space-time without black holes, we find $\omega _{\Lambda }^{\alpha
-DE}=\frac{\kappa }{3}+\sum_{k=2}^{K}\left(
\begin{array}{c}
K \\
k%
\end{array}%
\right) \frac{\alpha \left( \kappa \left( K\alpha \right) ^{\frac{1}{K-1}%
}\right) ^{k}H^{2k-2}}{3\left( K\alpha \right) ^{\frac{K}{K-1}}}.$ For a
space-time without Lovelock couplings, we find $\omega _{\Lambda }^{DE-BH}=%
\frac{\kappa }{3}-\frac{8G_{N}\Gamma \left( \frac{D-1}{2}\right) }{3\left(
D-2\right) \pi ^{\frac{\left( D-3\right) }{2}}}MH^{D-3}$. For a space-time
without black holes and Lovelock couplings, we find $\omega _{\Lambda }^{DE}=%
\frac{\kappa }{3}$. If $\kappa =-1$\ we find $\omega _{\Lambda }^{DE}=-\frac{%
1}{3}$, which is in good agreement with the EoS found in the framework of
the agegraphic DE \cite{L12}.

\section{Third-order Lovelock black holes}

Lovelock gravity is the most general p-order gravity theory in
higher-dimensional spacetimes. The action of the third-order Lovelock
gravity is given by \cite{Z1,Z2}:

\begin{equation}
\mathcal{I}=\frac{1}{16\pi }\int d^{D}x\sqrt{-g}\left( -2\Lambda +R+\alpha
_{2}\mathcal{L}_{2}+\alpha _{3}\mathcal{L}_{3}\right) ,  \label{c1}
\end{equation}%
where $\Lambda $ is the cosmological constant, $\alpha _{2}\mathcal{\ }$and $%
\alpha _{3}$ are respectively the second (Gauss-Bonnet coupling) and the
third Lovelock coefficients. Additionally, $\mathcal{L}_{1}=R$ is the
Einstein-Hilbert Lagrangian. The cosmological constant is $\Lambda =-\frac{%
\left( D-1\right) \left( D-2\right) }{2L^{2}},$ where $L$ is the AdS radius.
We introduce the effective thermodynamic pressure $P=-\Lambda /8\pi $ as
\begin{equation}
P=\frac{\left( D-1\right) \left( D-2\right) }{16\pi L^{2}}.  \label{cc}
\end{equation}%
The Gauss-Bonnet invariant and the third-order Lovelock invariant are
respectively%
\begin{equation}
\mathcal{L}_{2}=R^{2}-4R_{\mu \nu }R^{\mu \nu }+R_{\mu \nu \rho \sigma
}R^{\mu \nu \rho \sigma },  \label{c3}
\end{equation}%
\begin{eqnarray}
\mathcal{L}_{3} &=&R^{3}+2R^{\mu \nu \sigma k}R_{\sigma k\rho \tau }R_{\text{
\ }\mu \nu }^{\rho \tau }+8R_{\text{ \ \ }\sigma \rho }^{\mu \nu }R_{\text{
\ }\nu \tau }^{\sigma k}R_{\text{ \ }\mu k}^{\rho \tau }+24R^{\mu \nu \sigma
k}R_{\sigma k\nu \rho }R_{\text{ \ }\mu }^{\rho }  \label{c4} \\
&&+3RR^{\mu \nu \sigma k}R_{\mu \nu \sigma k}+24R^{\mu \nu \sigma
k}R_{\sigma \mu }R_{k\nu }+16R^{\mu \nu }R_{\nu \sigma }R_{\text{ \ }\mu
}^{\sigma }-12RR^{\mu \nu }R_{\mu \nu }.  \notag
\end{eqnarray}%
After the variation of the action with respect to the metric $g_{\mu \nu }$
\cite{T1}. We will use the following spherically symmetric ansatz for the
metric. As is shown in \cite{Z3}, the D-dimensional static solutions in
third order Lovelock gravity for $\alpha _{3}=\alpha ^{2}/72\left(
\begin{array}{c}
D-3 \\
4%
\end{array}%
\right) $ and $\alpha _{2}=\frac{\alpha }{\left( D-3\right) \left(
D-4\right) }$\ is given by the metric function as%
\begin{equation}
f(r)=\kappa +\frac{r^{2}}{\alpha }\left( 1-\left( 1+\frac{3\alpha m}{\left(
D-2\right) r^{D-1}}+\frac{6\alpha \Lambda }{\left( D-1\right) \left(
D-2\right) }\right) ^{1/3}\right) ,  \label{c6}
\end{equation}%
where $\alpha $ is the rescaled Lovelock coupling and $\kappa $ is the
horizon curvatures with $\kappa =\left\{ 0,\pm 1\right\} $\ for\ $\Lambda <0$%
\ and $\kappa =+1$\ for\ $\Lambda \geq 0$ and $m$ is an integration
constant. Thus we can uniformly rewrite the ADM mass $M$ \ for topological
black holes \cite{L15} as $M=\frac{V_{D-1}}{16\pi R_{H}^{D-1}}m.$ The $%
\left( D-1\right) $-dimensional volume of a Euclidean ball of radius $R_{H}$%
: \cite{CQG}:%
\begin{equation}
V_{D-1}=\frac{2\pi ^{\frac{\left( D-1\right) }{2}}}{\left( D-1\right) \Gamma
\left( \frac{D-1}{2}\right) }R_{H}^{D-1}.  \label{d2}
\end{equation}%
Here, $R_{H}$\ denotes the radius of the black hole and is one of the roots
of the metric function (\ref{c6}). We can express the ADM mass $M$ of the
black hole in terms of $R_{H}$:%
\begin{equation}
M=\frac{\left( D-2\right) R_{H}^{D-3}\pi ^{\frac{\left( D-3\right) }{2}}}{%
8\Gamma \left( \frac{D-1}{2}\right) }\left( \kappa +\frac{16\pi PR_{H}^{2}}{%
\left( D-1\right) \left( D-2\right) }+\frac{\alpha \kappa ^{2}}{R_{H}^{2}}+%
\frac{\alpha ^{2}\kappa }{3R_{H}^{4}}\right) .  \label{c10}
\end{equation}%
The Hawking temperature of the third-order Lovelock black hole can be
calculated as

\begin{equation}
T=\frac{1}{12\pi R_{H}\left( R_{H}^{2}+\kappa \alpha \right) ^{2}}\left(
\frac{48\pi R_{H}^{6}P}{\left( D-2\right) }+3\left( D-3\right)
R_{H}^{4}\kappa +3\left( D-5\right) R_{H}^{2}\alpha \kappa ^{2}+\left(
D-7\right) \alpha ^{2}\kappa \right) .  \label{c11}
\end{equation}%
In the limit $\alpha \longrightarrow 0$, we can recover $T=\frac{\left(
D-1\right) R_{H}}{4\pi L^{2}}+\frac{\left( D-3\right) \kappa }{4\pi R_{H}}.$
This limit corresponds to the standard Hawking temperature, if $D=4,$ $%
\kappa =+1$ and $L\rightarrow \infty $, i.e. $T=\frac{1}{4\pi R_{H}}$.
Consequently, the black hole entropy \cite{Z8} is
\begin{equation}
S=\frac{R_{H}^{D-2}}{4}\frac{2\pi ^{\frac{\left( D-1\right) }{2}}}{\Gamma
\left( \frac{D-1}{2}\right) }\left( 1+\frac{2\left( D-2\right) \kappa \alpha
}{\left( D-4\right) R_{H}^{2}}+\frac{\left( D-2\right) \kappa ^{2}\alpha ^{2}%
}{\left( D-6\right) R_{H}^{4}}\right) .  \label{c12}
\end{equation}%
In the limit $\alpha \longrightarrow 0$, one can recover the
Bekenstein-Hawking entropy in $D=4$.

\section{Energy conditions from black hole mass}

Next, we study the different densities associated with black hole mass and
energy conditions. Also, we give physical interpretations to the rescaled
Lovelock coupling. From Eq. (\ref{c10}) and using $\frac{P}{\left(
D-2\right) \left( D-1\right) }=\frac{1}{16\pi L^{2}}$, we obtain%
\begin{equation}
\frac{3}{16\pi L^{2}}+\frac{3\kappa }{16\pi R_{H}^{2}}+\frac{3\alpha \kappa
^{2}}{16\pi R_{H}^{4}}+\frac{\alpha ^{2}\kappa }{16\pi R_{H}^{6}}=\frac{3M}{%
\left( D-2\right) \left( D-1\right) V_{D-1}}.  \label{M1}
\end{equation}%
In the limit $\alpha \longrightarrow 0$, we can recover the density of the
holographic DE \cite{D9}: $\rho _{\Lambda }=\frac{3}{16\pi R_{H}^{2}}.$ For
the second and third-order Lovelock branches Eq. (\ref{M1}), we introduce
the topological densities as%
\begin{equation}
\rho _{\alpha }^{\left( 0\right) }=\frac{3}{16\pi L^{2}},\text{ }\rho
_{\alpha }^{\left( 1\right) }=\frac{3}{16\pi R_{H}^{2}},\text{ }\rho
_{\alpha }^{\left( 2\right) }=\frac{3\alpha }{16\pi R_{H}^{4}},\text{ }\rho
_{\alpha }^{\left( 3\right) }=\frac{3\alpha ^{2}}{16\pi R_{H}^{6}},
\label{C0}
\end{equation}%
with $\rho _{\alpha }^{\left( 0\right) }=\frac{3}{\left( D-2\right) \left(
D-1\right) }P$ and $\rho _{\alpha }^{\left( 1\right) }=\rho _{\Lambda }$. It
is easy to check that Eq. (\ref{C0}) satisfies: $\rho _{\alpha }^{\left(
0\right) }=\frac{3}{16\pi L^{2}}$ and $\rho _{\alpha }^{\left( n\right) }=%
\frac{3\alpha ^{n-1}}{16\pi R_{H}^{2n}},$ $1\leq n\leq 3$. From this, it is
clear that the Lovelock coupling is similar to a topological mass. We see
that $\rho _{\alpha }^{\left( n\right) }\propto \alpha ^{n-1}\rho _{\Lambda
} $, which shows that the black hole densities $\rho _{\alpha }^{\left(
n\right) }$ in second-third order Lovelock gravity depend on the dark energy
density $\rho _{\Lambda }$\textrm{.} The general case of the density
equation can thus be written as
\begin{equation}
\frac{3P}{\left( D-2\right) \left( D-1\right) }+\kappa \rho _{\Lambda
}+\kappa ^{2}\rho _{\alpha }^{\left( 2\right) }+\frac{1}{3}\kappa \rho
_{\alpha }^{\left( 3\right) }=\frac{3M}{\left( D-2\right) \left( D-1\right)
V_{D-1}}.  \label{M2}
\end{equation}%
or equivalently%
\begin{equation}
\frac{3\left( \rho _{BH}-P\right) }{\left( D-2\right) \left( D-1\right) }%
=\kappa \rho _{\Lambda }+\kappa ^{2}\rho _{\alpha }^{\left( 2\right) }+\frac{%
1}{3}\kappa \rho _{\alpha }^{\left( 3\right) },  \label{mw}
\end{equation}%
with $\rho _{BH}=\frac{M}{V_{D-1}}>0$. The solution with flat horizon $%
\kappa =0$\ in the above expressions corresponds to the ADM mass $M=PV_{D-1}$%
\ or equivalently $P=\rho _{BH}$. Using topological/geometrical properties
of spacetime and the energy conditions to build a new type of DE. In order
to check the viability of holographic dark energy with black holes in
Lovelock gravity, let us discuss the energy conditions \cite{L16}. We start
first with the system $\left( P,\rho _{BH}\right) $.\textbf{\ }The
relationship (\ref{mw}) verified the DEC condition $\rho _{BH}>0$ and $\rho
_{BH}\geq \left\vert P\right\vert $. It is easy to check that $\rho
_{BH}\geq 0$. This satisfies the strong energy condition $\rho _{BH}+3P\geq
0 $. One can also see that $\rho _{BH}+P\geq 0$, which satisfied the null
energy condition. In concluding, the presence of a black hole in the
pressure $P$ satisfies all the energy conditions. By considering a specific
pressure of solutions corresponding to the choice of
\begin{equation}
\tilde{P}=\frac{3P}{\left( D-2\right) \left( D-1\right) }\equiv -\omega
_{\Lambda }\rho _{\Lambda },  \label{M3}
\end{equation}%
with $\omega _{\Lambda }$ is the EoS parameter of dark energy. This pressure
does not depend on the number of space-time dimensions. So that one obtains
the following solution%
\begin{equation}
\tilde{P}+\kappa \rho _{\Lambda }+\kappa ^{2}\rho _{\alpha }^{\left(
2\right) }+\frac{1}{3}\kappa \rho _{\alpha }^{\left( 3\right) }=\frac{3\rho
_{BH}}{\left( D-2\right) \left( D-1\right) }.  \label{M4}
\end{equation}%
This relationship describes the connection between the black holes and
holographic DE \cite{L13}. Here, we analyze the energy conditions for the
above equation.\newline
$\bullet $ We consider the system $\left( \tilde{P},\rho _{\Lambda }\right) $
and we take $\kappa =+1$ (for $\Lambda <0$ or $\Lambda \geq 0$): In this
case, Eq .(\ref{M4}) take the form $\tilde{P}+\rho _{\Lambda }+\rho _{\alpha
}^{\left( 2\right) }+\frac{1}{3}\rho _{\alpha }^{\left( 3\right) }=\frac{%
3\rho _{BH}}{\left( D-2\right) \left( D-1\right) }$. The NEC condition ($%
\tilde{P}+\rho _{\Lambda }\geq 0$) and the WEC condition ($\tilde{P}+\rho
_{\Lambda }\geq 0$ and $\rho _{\Lambda }\geq 0$) reduces to $3\rho _{\alpha
}^{\left( 2\right) }+\rho _{\alpha }^{\left( 3\right) }\leq \frac{9\rho _{BH}%
}{\left( D-1\right) \left( D-2\right) }$ and $\rho _{\Lambda }\geq 0$. For $%
\kappa =-1$ ($\Lambda <0)$, the Eq. (\ref{M4}) reduce to $\rho _{\Lambda }-%
\tilde{P}=\rho _{\alpha }^{\left( 2\right) }-\frac{3\rho _{BH}}{\left(
D-1\right) \left( D-2\right) }-\frac{\rho _{\alpha }^{\left( 3\right) }}{3}$%
, which may be satisfied for DEC condition ($\rho _{\Lambda }\geq \left\vert
\tilde{P}\right\vert $ and $\rho _{\Lambda }\geq 0$) as $\rho _{\alpha
}^{\left( 2\right) }-\frac{\rho _{\alpha }^{\left( 3\right) }}{3}\geq \frac{%
3\rho _{BH}}{\left( D-1\right) \left( D-2\right) }$.\newline
$\bullet $ We consider the system $\left( P,\rho _{\alpha }^{\left( 3\right)
}\right) $ and we take $\kappa =+1$ Eq .(\ref{M4}): In this case, the SEC
condition ($\tilde{P}+\rho _{\alpha }^{\left( 3\right) }\geq 0$ and $\left(
D-1\right) \tilde{P}+\left( D-3\right) \rho _{\alpha }^{\left( 3\right)
}\geq 0$) is satisfied if $D=4$\ and $\rho _{\Lambda }+\rho _{\alpha
}^{\left( 2\right) }\leq \frac{3\rho _{BH}}{\left( D-1\right) \left(
D-2\right) }$. In concluding, for the asymptotically flat case, the NEC,
WEC, DEC and SEC are satisfied with a suitable choice of parameters. Taking $%
L=H^{-1}$ as the effective Hubble horizon, thus the Friedmann equation $%
HR_{H}=1$ \cite{D9}, yielding
\begin{equation}
\omega _{\Lambda }=\kappa -\frac{8\Gamma \left( \frac{D-1}{2}\right) }{\pi ^{%
\frac{\left( D-3\right) }{2}}\left( D-2\right) }MH^{D-3}+\kappa ^{2}\alpha
H^{2}+\frac{\kappa \alpha ^{2}}{3}H^{4}.  \label{og}
\end{equation}%
By taking into account\textrm{\ }$\alpha =0$ and $\kappa =1$ one arrives at
the following EoS $\omega _{\Lambda }=1-2MH$ for $D=3$ and $\Gamma \left(
\frac{3}{2}\right) =\frac{\sqrt{\pi }}{2}$. From this, it is clear that
satisfying the non-phantom dark energy if $MH\leq 1$ or equivalently $\omega
_{\Lambda }\geq -1$ \cite{KKR}. The DE-EoS of this model depends on both
Hubble parameter $H$ and the rescaled Lovelock coupling. If we consider a
space-time without a black hole, we find $\omega _{\Lambda }=\kappa +\kappa
^{2}\alpha H^{2}+\frac{\kappa \alpha ^{2}}{3}H^{4}$. In particular, we
consider a space-time without black holes ($M=0$) and the topological nature
($\alpha =0$), one obtains $\omega _{\Lambda }=-1$ for $\kappa =-1$, which
represents the EoS of dark energy. The radiations evolve in such a way $%
H^{2}\sim T^{4}\propto \omega _{\Lambda }$, which is in good agreement with
\cite{N2}.

\section{Energy conditions from black hole temperature}

In what follows, we aim to present the solutions with the energy conditions
using the Hawking temperature in third-order Lovelock gravity. From Eq. (\ref%
{c11}) and using $\frac{P}{\left( D-2\right) }=\frac{\left( D-1\right) }{%
16\pi L^{2}}$ we obtain%
\begin{equation}
\frac{P}{\left( D-2\right) }+\left( D-3\right) \kappa \frac{1}{16\pi
R_{H}^{2}}+\left( D-5\right) \kappa ^{2}\frac{\alpha }{16\pi R_{H}^{4}}%
+\left( D-7\right) \frac{\kappa }{3}\frac{\alpha ^{2}}{16\pi R_{H}^{6}}%
=\left( R_{H}^{2}+\kappa \alpha \right) ^{2}\frac{T}{4R_{H}^{5}}.  \label{T2}
\end{equation}%
Combining Eqs. (\ref{C0}), (\ref{M3}) and (\ref{T2}) we deduce that%
\begin{equation}
\left( 1+\frac{\kappa \alpha }{R_{H}^{2}}\right) ^{2}\frac{3T}{4R_{H}}%
=\left( D-1\right) \tilde{P}+\left( D-3\right) \kappa \rho _{\Lambda
}+\left( D-5\right) \kappa ^{2}\rho _{\alpha }^{\left( 2\right) }+\left(
D-7\right) \frac{1}{3}\kappa \rho _{\alpha }^{\left( 3\right) }.  \label{T3}
\end{equation}%
In order to check the viability of third-order Lovelock densities, let us
discuss the energy condition. First, we consider the system $\left( P,\rho
_{\Lambda }\right) $ and we take $\kappa =+1$ (for $\Lambda <0$ or $\Lambda
\geq 0$): In this case, It is easy to check that Eq .(\ref{T3}) satisfies
the SEC, NEC and WEC conditions ($\tilde{P}+\rho _{\Lambda }\geq 0$, $\rho
_{\Lambda }\geq 0$ and $\left( D-1\right) \tilde{P}+\left( D-3\right) \rho
_{\Lambda }\geq 0$) if%
\begin{equation}
\left( 1+\frac{\kappa \alpha }{R_{H}^{2}}\right) ^{2}\frac{3T}{4R_{H}}\geq
\left( D-5\right) \kappa ^{2}\rho _{\alpha }^{\left( 2\right) }+\left(
D-7\right) \frac{1}{3}\kappa \rho _{\alpha }^{\left( 3\right) },
\end{equation}%
\begin{equation}
3\tilde{P}+\rho _{\Lambda }+\left( 1+\frac{\kappa \alpha }{R_{H}^{2}}\right)
^{2}\frac{3T}{4R_{H}}\geq \left( D-5\right) \rho _{\alpha }^{\left( 2\right)
}+\left( D-7\right) \frac{1}{3}\rho _{\alpha }^{\left( 3\right) }.
\end{equation}%
Obviously when $\alpha =0$, we obtain a condition on the temperature: $3%
\tilde{P}+\rho _{\Lambda }+\frac{3T}{4R_{H}}\geq 0$, which is related to the
SEC condition if $D=4$. For $\kappa =-1$ ($\Lambda <0)$, it is easy to check
that Eq .(\ref{T3}) satisfies the DEC ($\rho _{\Lambda }\geq \left\vert
\tilde{P}\right\vert $ and $\rho _{\Lambda }\geq 0$) as%
\begin{equation}
\rho _{\Lambda }-3\tilde{P}+\left( D-5\right) \rho _{\alpha }^{\left(
2\right) }+\left( D-7\right) \frac{1}{3}\rho _{\alpha }^{\left( 3\right)
}\geq \left( 1-\frac{\alpha }{R_{H}^{2}}\right) ^{2}\frac{3T}{4R_{H}}.
\end{equation}%
When $\alpha =0$, we obtain a condition on the temperature $\rho _{\Lambda
}-3\tilde{P}\geq \frac{3T}{4R_{H}}$. Using $HR_{H}=1$ \cite{D9} for the
general case and by taking into account Eq. (\ref{T3}), one arrives at the
following form of DE EoS parameter:%
\begin{equation}
\omega _{\Lambda }=\frac{\left( D-3\right) }{\left( D-1\right) }\kappa
-\left( 1+\kappa \alpha H^{2}\right) ^{2}\frac{4\pi T}{\left( D-1\right) H}+%
\frac{\left( D-5\right) }{\left( D-1\right) }\alpha \kappa ^{2}H^{2}+\frac{%
\left( D-7\right) }{\left( D-1\right) }\frac{\alpha ^{2}\kappa }{3}H^{4}.
\label{op}
\end{equation}%
One can see from this equation that $\omega _{\Lambda }=-\frac{4\pi T}{%
\left( D-1\right) H}$ for $\kappa =0$. Obviously, we have $4\pi T\leq \left(
D-1\right) H$ reduces to $\omega _{\Lambda }\geq -1$. Therefore this EoS
describes non-phantom dark energy. It is worthwhile to mention that due to
the absence of the temperature and for $\alpha =0$, we have $\omega
_{\Lambda }=\kappa \frac{\left( D-3\right) }{\left( D-1\right) }$, in $D=4$
one obtain $\omega _{\Lambda }=\frac{\kappa }{3}$. If $\kappa =-1$, we have $%
\omega _{\Lambda }=-\frac{1}{3}$. Note that for $\kappa =1$ the effective
EoS parameter at the critical point is $\omega _{\Lambda }=\frac{1}{3}$ \cite%
{T2}, which means that its energy density decreases as $a^{-4}$. In the
limit $\alpha \longrightarrow 0$, we can recover the effective EoS parameter
$\omega _{\Lambda }\approx -\frac{1}{3}+\frac{4\pi T}{3H}$ in $D=4$ with $%
\kappa =-1$. In comparing this result with those outlined in\ the origin of
the flatness problems \cite{T3}, we see that the EoS parameter ($\omega
_{\Lambda }\approx -\frac{1}{3}$) for expanding universe describes this
problem if $T=0$. In concluding, the energy condition of the 4D black hole
temperature with $T=0$ presents the problem of flatness. From Eq. (\ref{c12}%
), (\ref{C0}) and (\ref{T2}) the resulting density and for $\kappa =0$ ($%
\Lambda <0$) we deduce that $S=\frac{4\pi R_{H}V_{D-1}}{3}\rho _{\Lambda }.$
For an ideal gas ($\tilde{P}V_{D-1}=\tilde{T}$), we have $S\propto \tilde{T}%
\left\vert \omega _{\Lambda }\right\vert $. The existence of dark energy in
the black hole horizon can be investigated from this entropy.
\begin{figure}[H]
\centering\includegraphics[width=11cm]{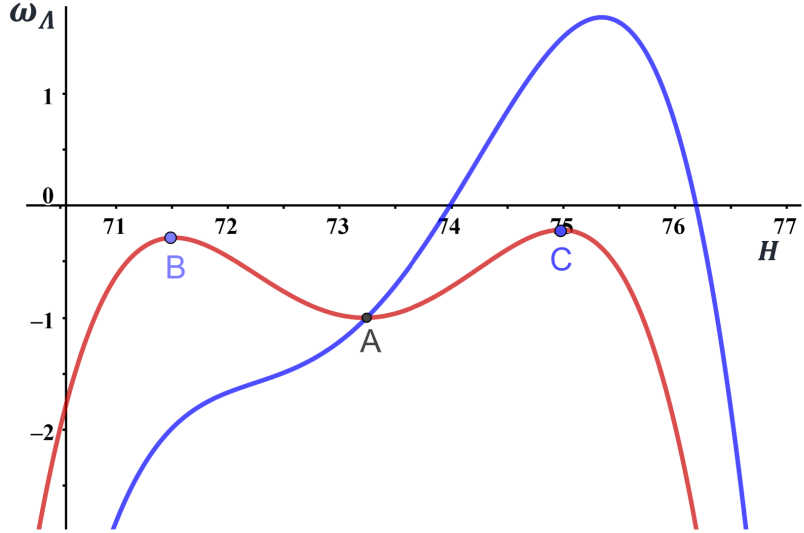}
\caption{Curves of $\protect\omega _{\Lambda }\left( H-73.24\right) $ versus
$H$ in $D=4$ for $k=-1$ and $\protect\alpha =0.5$\ from Eq.\ (\protect\ref%
{og}) for $H\rightarrow H-73.24$. The blue and red curves are determined
respectively for $M=0.5$ and $M=0.01$. The points A, B and C correspond
respectively to the following values: $\left[ H_{A}=73.24;\text{ }\protect%
\omega _{\Lambda }(A)=-1\right] ,$ $\left[ H_{B}=71.48;\text{ }\protect%
\omega _{\Lambda }(B)=-0.29\right] $\ and $\left[ H_{C}=74.99;\text{ }%
\protect\omega _{\Lambda }(C)=-0.22\right] .$}
\label{F1}
\end{figure}
\begin{figure}[H]
\centering\includegraphics[width=11cm]{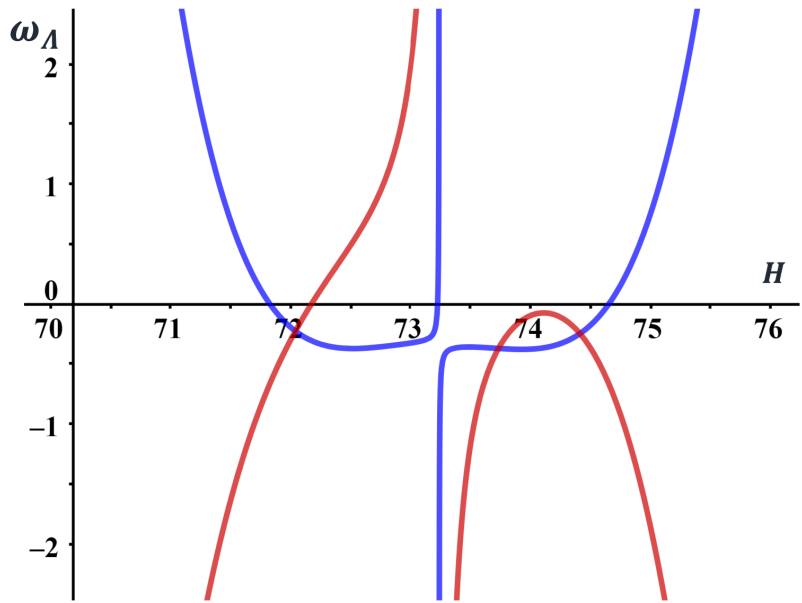}
\caption{Curves of $\protect\omega _{\Lambda }\left( H-73.24\right) $ versus
$H$\ in $D=4$ from Eq.\ (\protect\ref{op}) for $H\rightarrow H-73.24$ and $%
\protect\alpha =0.5$. The blue and red curves are determined respectively by$%
\left( \protect\kappa =-1,\text{ }T=0.001\right) $ and $\left( \protect%
\kappa =+1,\text{ }T=0.1\right) $. }
\label{F2}
\end{figure}
Using the numerical analysis of Eqs. (\ref{og})-(\ref{op}) as shown in Figs.
(\ref{F1})-(\ref{F2}). We observe that Fig. (\ref{F1}) describes a realistic
case of the dark energy with $\omega _{\Lambda }=-1$ and the Hubble constant
$H=73.24$ km s$^{-1}$ Mpc$^{-1}$ \cite{L1}. Point A represents a stable
equilibrium of $\omega _{\Lambda }$. While points B and C describe are
unstable equilibrium of $\omega _{\Lambda }$. Fig. (\ref{F2}) shows a broad
domain of phantom dark energy, while, the minimum values of the blue curve
describe non-phantom states.

\section{Conclusion}

In this paper, we have demonstrated the energy conditions for Lovelock black
hole searches for dark energy, showing the equation of state parameter for
dark energy. Inspired by the energy density \cite{D9} based on holographic
principle \cite{D11,D12}. The main motivation behind the satisfaction of the
energy conditions is to find the link between the variation of the
geometrical/topological parameters and the energy parameters. The geometric
parameters locally describe the energy, for example dark energy. On the
other hand, the topological parameter is a global description of the black
hole. We have obtained the analytic solutions for dark energy and
topological densities by using black hole mass and temperature. We have
shown that the parameter $\omega _{\Lambda }$\ of this model depends on both
Hubble parameter $H$\ (with $H=R_{H}^{-1}$\ \cite{D9}) and the rescaled
Lovelock coupling\ which is governed by horizon curvatures. We have also
included the possibility of having unstable and stable equilibriums of $%
\omega _{\Lambda }$. \newline
Additionally, the Lovelock gravity admits cosmological solutions of the
holographic dark energy. This is a consequence of topological/geometrical
properties of spacetime, which made it possible to study accelerated
expansion of the Universe by using the results of the Lovelock black hole.
As we have discussed, however, there is a connection between black hole and
holographic dark energy \cite{L13}. This provides an alternative way to
explain cosmic acceleration with no need of invoking the Friedmann equations
but within the context of black holes. Future work will have to test if this
model also corresponds to the CMB observations.\newline
We thank Neda Farhangkhah for useful conversations. We thank Sunny Vagnozzi
and Liu Zhao for detailed comments.\newline

\section{Appendix}

1- Detail on Eq. (\ref{M1}):

From Eq. (\ref{c10}) we have

\begin{equation*}
M=\frac{\left( D-2\right) R_{H}^{D-3}\pi ^{\frac{\left( D-3\right) }{2}}}{%
8\Gamma \left( \frac{D-1}{2}\right) }\left( \kappa +\frac{16\pi PR_{H}^{2}}{%
\left( D-1\right) \left( D-2\right) }+\frac{\alpha \kappa ^{2}}{R_{H}^{2}}+%
\frac{\alpha ^{2}\kappa }{3R_{H}^{4}}\right) .
\end{equation*}%
\begin{equation*}
\frac{3\kappa }{16\pi }+\frac{3PR_{H}^{2}}{\left( D-1\right) \left(
D-2\right) }+\frac{3\alpha \kappa ^{2}}{16\pi R_{H}^{2}}+\frac{\alpha
^{2}\kappa }{16\pi R_{H}^{4}}=\frac{8\Gamma \left( \frac{D-1}{2}\right) }{%
16\pi \left( D-2\right) R_{H}^{D-3}\pi ^{\frac{\left( D-3\right) }{2}}}3M.
\end{equation*}%
\begin{equation*}
\frac{3\kappa }{16\pi }+\frac{3PR_{H}^{2}}{\left( D-1\right) \left(
D-2\right) }+\frac{3\alpha \kappa ^{2}}{16\pi R_{H}^{2}}+\frac{\alpha
^{2}\kappa }{16\pi R_{H}^{4}}=\frac{\left( D-1\right) \Gamma \left( \frac{D-1%
}{2}\right) }{2\pi ^{\frac{\left( D-1\right) }{2}}R_{H}^{D-1}}\frac{%
3MR_{H}^{2}}{\left( D-1\right) \left( D-2\right) }.
\end{equation*}%
\begin{equation*}
\frac{3\kappa }{16\pi R_{H}^{2}}+\frac{3PR_{H}^{2}}{\left( D-1\right) \left(
D-2\right) R_{H}^{2}}+\frac{3\alpha \kappa ^{2}}{16\pi R_{H}^{4}}+\frac{%
\alpha ^{2}\kappa }{16\pi R_{H}^{6}}=\frac{\left( D-1\right) \Gamma \left(
\frac{D-1}{2}\right) }{2\pi ^{\frac{\left( D-1\right) }{2}}R_{H}^{D-1}}\frac{%
3M}{\left( D-1\right) \left( D-2\right) }.
\end{equation*}%
Using Eq. (\ref{d2}), we find%
\begin{equation*}
\frac{3}{16\pi L^{2}}+\frac{3\kappa }{16\pi R_{H}^{2}}+\frac{3\alpha \kappa
^{2}}{16\pi R_{H}^{4}}+\frac{\alpha ^{2}\kappa }{16\pi R_{H}^{6}}=\frac{3M}{%
\left( D-2\right) \left( D-1\right) V_{D-1}}.
\end{equation*}%
2- Detail on Eq. (\ref{T3}):\newline
From Eq. (\ref{c11}) we have%
\begin{equation}
T=\frac{1}{12\pi R_{H}\left( R_{H}^{2}+\kappa \alpha \right) ^{2}}\left(
\frac{48\pi R_{H}^{6}P}{\left( D-2\right) }+3\left( D-3\right)
R_{H}^{4}\kappa +3\left( D-5\right) R_{H}^{2}\alpha \kappa ^{2}+\left(
D-7\right) \alpha ^{2}\kappa \right) .
\end{equation}%
\begin{equation}
\frac{48\pi R_{H}^{6}P}{\left( D-2\right) }+3\left( D-3\right)
R_{H}^{4}\kappa +3\left( D-5\right) R_{H}^{2}\alpha \kappa ^{2}+\left(
D-7\right) \alpha ^{2}\kappa =12\pi R_{H}\left( R_{H}^{2}+\kappa \alpha
\right) ^{2}T.
\end{equation}%
\begin{equation}
\frac{P}{\left( D-2\right) }+\left( D-3\right) \frac{3R_{H}^{4}}{48\pi
R_{H}^{6}}\kappa +3\left( D-5\right) \frac{R_{H}^{2}\alpha }{48\pi R_{H}^{6}}%
\kappa ^{2}+\left( D-7\right) \frac{\alpha ^{2}}{48\pi R_{H}^{6}}\kappa =%
\frac{12\pi R_{H}}{48\pi R_{H}^{6}}\left( R_{H}^{2}+\kappa \alpha \right)
^{2}T.
\end{equation}%
we obtain%
\begin{equation}
\frac{P}{\left( D-2\right) }+\left( D-3\right) \kappa \frac{1}{16\pi
R_{H}^{2}}+\left( D-5\right) \kappa ^{2}\frac{\alpha }{16\pi R_{H}^{4}}%
+\left( D-7\right) \frac{\kappa }{3}\frac{\alpha ^{2}}{16\pi R_{H}^{6}}%
=\left( R_{H}^{2}+\kappa \alpha \right) ^{2}\frac{T}{4R_{H}^{5}}.
\end{equation}%
Combining Eqs. (\ref{C0}), (\ref{M3}) and (\ref{T2}) we deduce that%
\begin{equation}
\left( D-1\right) \tilde{P}+\left( D-3\right) \kappa \frac{3}{16\pi R_{H}^{2}%
}+\left( D-5\right) \kappa ^{2}\frac{3\alpha }{16\pi R_{H}^{4}}+\left(
D-7\right) \kappa \frac{\alpha ^{2}}{16\pi R_{H}^{6}}=\left(
R_{H}^{2}+\kappa \alpha \right) ^{2}\frac{3T}{4R_{H}^{5}}.
\end{equation}%
\begin{equation}
\left( 1+\frac{\kappa \alpha }{R_{H}^{2}}\right) ^{2}\frac{3T}{4R_{H}}%
=\left( D-1\right) \tilde{P}+\left( D-3\right) \kappa \rho _{\Lambda
}+\left( D-5\right) \kappa ^{2}\rho _{\alpha }^{\left( 2\right) }+\left(
D-7\right) \frac{1}{3}\kappa \rho _{\alpha }^{\left( 3\right) }.
\end{equation}

\end{document}